\documentclass[12pt]{article}

\newcommand{\be}{\begin{equation}}
\newcommand{\ee}{\end{equation}}
\newcommand{\bea}{\begin{eqnarray}}
\newcommand{\eea}{\end{eqnarray}}
\newcommand{\nn}{\nonumber}
\newcommand{\pt}{\stackrel{\leftarrow}{\partial_t}{}\!\!}
\newcommand{\gP}{\bar\mathcal{P}}
\newcommand{\Pg}{\mathcal{P}}
\newcommand{\half}{\frac{1}{2}}

\begin{document}
\begin{titlepage}


\title{Canonical approach to Lagrange multipliers}

\author{M. Stoilov\\
    Institute of Nuclear Research and Nuclear Energy,
        Sofia 1784, Bulgaria\\
e-mail: {\rm mstoilov@inrne.bas.bg}}

\date{2 May 2005}
\maketitle


\begin{abstract}
Lagrange multipliers are present in any gauge theory.
They possess peculiar gauge transformation which is
not generated by the constraints in the model as it is the case with 
the other variables.
For rank one gauge theories
we show how to alter the constraints so that they become 
generators of the local symmetry algebra in the space of
 Lagrange multipliers too.
We also discuss the limitations on using different gauge conditions and
construct the BRST charge corresponding to the newly found constraints.
\end{abstract}

\thispagestyle{empty}
\end{titlepage}

Gauge theories are essential part of contemporary physics.
There is a vast literature on the subject; one could become
familiar with the  problem from, e.g. refs. \cite{FS}, \cite{Hen}.
Gauge theories are best understood in the Hamiltonian approach, or equivalently, 
first order Lagrangian formalism with Lagrangian
\footnote{We shall not deal with the question how the Lagrangian (\ref{lag})
is obtained from the corresponding second order Lagrangian $L'(q,\dot q)$. 
For more information on the subject see, e.g. \cite{GT}.}
\be
L=p\dot q - H - \lambda_a \varphi_a \label{lag}
\ee
An implicit summation over all degrees of freedom 
 (which could be discrete as well as continuance) is understood. 
Here $H$ is the Hamiltonian of the system, $\lambda_a$ are the
Lagrange multipliers and $\varphi_a$ are constraints. 
Constraints are some (independent) functions of 
phase space variables $\{q, p\}$ which we assume to be first class  
(see below eqs.(\ref{alge}) for definition).
In this case 
the dynamics and gauge transformation of any dynamical quantity $g$
are governed by, respectively, the Hamiltonian and  
constraints in the model via Poisson brackets
\bea
\dot g &=&\{g,H\}_{PB}\nn\\
\delta_\epsilon g &=& \{g, \epsilon_a \varphi_a\}_{PB}. \label{dyn}
\eea
Here $\epsilon_a$ are arbitrary gauge parameters.
The only exception of this general rule are Lagrangian multipliers ---
neither time evolution nor gauge transformation of $\lambda_a$ are
determined by eqs.(\ref{dyn}).
In the present article we would like to modify $H$ and $\varphi_a$ in order
to allow uniform treating of these multipliers too, i.e.,
we want to find equations like (\ref{dyn}) for $\dot\lambda_a$ and 
$\delta_\epsilon \lambda_a$.
(For a different viewpoint and an alternative approach using Noether symmetries 
see \cite{DE},\cite{PG} and references therein.)
It turns out that this modification is possible only for the constraints,
while the time dependence of $\lambda_a$ remains completely undetermined. 
The second question we want to address here is about gauge fixing --- 
it is possible to fix the gauge either by imposing 
conditions of the type $\chi_a(q,p)=0$ or by fixing the Lagrange multipliers. 
We want to investigate the relation between these two gauge conditions
and, eventually, to show their equivalece.
Finally, we construct the BRST charge for the modified constraints and
discuss its connection with the BRST charge from the BFV approach 
\cite{Hen},\cite{FL} .

In what follows we shall deal with rank one theories only.
Together with the fact that constraints are  first class this  
means that $\varphi_a$ satisfy the following Poisson bracket relations:
\bea
 \{\varphi_a, \varphi_b\}_{PB} & = &C_{abc}\varphi_c, \nn\\
\{H, \varphi_a\}_{PB} & = & U_{ab}\varphi_b, \label{alge}
\eea
where $U_{ab}$ and $C_{abc}$ do not depend on dynamical variables. 
Most of our  considerations below could be easily generalized to 
higher rank theories. 
In the latter case procedure resembles the construction of BRST charge
for such theories.
 
The Lagrangian (\ref{lag}) is invariant under gauge transformations 
generated by $\varphi_a$ provided that $\lambda_a$ transform as follows:
\be
\delta_\epsilon \lambda_a = \dot\epsilon_a -\epsilon_b U_{ba} +
\epsilon_c C_{cba}\lambda_b. \label{varlm}
\ee
En route to deriving eqs.(\ref{varlm}) we encounter the calculation of 
 $\delta_\epsilon \dot q$.
The formula we use is
\be
\delta_\epsilon \dot q = \dot{(\delta_\epsilon q)}.\label{dv}
\ee
It can be obtained using the equations of motion, or, in other 
words, we temporarily switch to second order Lagrangian description of the
theory, find the desired variarion and then go back to the Lagrangian (\ref{lag}).
Note that eq. (\ref{dv}) does not imply $U_{ab}=0$.

Our first aim is to obtain the gauge transformation (\ref{varlm})
in the same way as we get the
gauge transformation of any other dynamical quantity, namely, 
generated by constraints via Poisson brackets.
We are looking for $\hat\varphi_a$, such that
\be
\delta_\epsilon \lambda = \{\lambda, \epsilon_a\hat\varphi_a\}_{PB}.
\ee
Here $\hat\varphi_a=\varphi_a + \{$ terms involving $\lambda_b$ and 
their momenta $\pi_b \}$.
 Once we step on this way we have to modify all three terms in
the  Lagrangian (\ref{lag}) with a new Hamiltonian 
$\hat H = H + h(\lambda,\pi)$.
However, we do not want to modify neither the dynamics nor the gauge 
freedom in the theory.
This means that we want $\hat H$ and $\hat\varphi_a$ to satisfy
an algebra like (\ref{alge}) and eqs.(\ref{dyn}) to still hold.
 Therefor, we impose:
\bea
 \{\hat\varphi_a, \hat\varphi_b\}_{PB}&=&C_{abc}\hat\varphi_c,\nn\\
\{\hat H, \hat\varphi_a\}_{PB}&=&U_{ab}\hat\varphi_b,\nn\\
\{f(q,p), \hat\varphi_a\}_{PB} &=& \{f(q,p), \varphi_a\}_{PB}\label{fullalg}
\eea
for any function $f(q,p)$.
Using the Jacobi identities among the structure constants $C_{abc}$ and between
them and $U_{ab}$ we find the following expression for $\hat\varphi_a$:
\be
\hat\varphi_a=\varphi_a + \pt\pi_a - U_{ab} \pi_b + 
\lambda_b C_{abc}\pi_c. \label{tot}
\ee
The time derivative in the second term in the r.h.s. of the above expression
acts on the gauge parameters. 
Thus we can freely calculate Poisson brackets involving $\hat\varphi$.
Note that we need the Hamiltonian in order to construct the constraints 
(\ref{tot}).
This is not the case when one "prolongs" the constraints with ghost terms. 
Having $\hat\varphi_a$ we can find $\hat H$:
\be
\hat H=H + \lambda_a U_{ab} \pi_b.
\ee
Finally, in order for the total Lagrangian to be invariant under gauge 
transformations we have to add to it the term 
\be
\pi_a\dot\lambda_a
\ee
Putting all things together we get that all extra terms cancel out and 
\be
\hat L = L
\ee
As a result the dynamics of the Lagrangian multipliers is not determined and 
they are completely arbitrary as they should be.

Before constructing the BRST charge corresponding to the constraints (\ref{tot})
we shall switch our attention to the different possible gauge conditions. 
The straightforward approach is to pick up some functions
\be
\chi_a(q,p)=0 \label{gc}
\ee 
(coinciding in number with the constraints) such that 
\bea
\{\chi_a, \chi_b\}_{PB}&=&0, \nn\\
det \vert\{\chi)_a, \varphi_b\}_{PB}\vert &\ne& 0.
\eea
These conditions ensure that, first, we can choose $\chi_a$
as part of the configuration space variables and, second, that we can 
resolve constraints with respect to the  momenta corresponding to $\chi_a$.
As a result we obtain a description of the 
constrained system entirely in terms of independent phase space variables .
 In addition, as a consequence of the second 
of the above equations, there is no residual gauge freedom. 
Now we want to understand if the gauge conditions (\ref{gc})
fix the Lagrange multipliers.
Denoting  $\{\chi_a, \varphi_b\}_{PB}$ with $\Delta_{ab}$ 
and $\{\chi_a, H\}_{PB}$ with $\mu_a$ and
using the equations of motion for $\chi_a$ we obtain
\be
0=\dot\chi_a=\{\chi_a, H + \lambda_b\varphi_b\}_{PB}=
\mu_a + \Delta_{ab} \lambda_b,
\ee
and so, we find 
\be
\lambda=\Delta^{-1}\mu. \label{l1}
\ee
Thus, in general, imposing (\ref{gc}) we find $\lambda_a$. 
However, at this point we encounter a problem.
It is best manifested when 
we deal with zero canonical Hamiltonian (as in the string models for instance). 
In this case the unique solution of eqs.(\ref{l1}) is 
$\lambda_a=0$ and we end up with void theory. 
To understand this phenomenon we focus 
on using fixing of Lagrangian multipliers as gauge conditions. 
Suppose we set the Lagrangian multipliers to some constants $c_a$
\be
\lambda_a=c_a. \label{lc}
\ee 
As seen from eqs.(\ref{varlm}) this does not fix 
the gauge entirely --- we can freely make transformation with parameter
$$\epsilon = e^{-(U+C\lambda) t} \zeta.$$
In the case of field theory $\zeta_a$ are arbitrary functions of
the spatial coordinates.
This is a huge residual freedom in striking contrast to the situation 
when we use (\ref{gc}) as gauge conditions. 
Let us see if eqs.(\ref{lc}) are enough 
 to determine the physical degrees of freedom. 
In order to do this we recall that the momenta $p$ are introduced  
through Legandre transformation of $\dot q$. 
In the case of constrained theory this transformation is degenerate
and there is no one--to--one correspondence between 
$\dot q$ and $p$. 
However, there is such correspondence between $\dot q$ on the one 
side and $\lambda$ and independent $p$ on the other. 
Roughly speaking, fixing $\lambda$ one fixes not part of 
the dynamical coordinates $q$ but $\dot q$. 
To make this consideration more quantitative 
we need to exploit a procedure known as "Abelization" of the constraints. 
It is always possible (locally) to achieve Abelization with a canonical
transformation. 
For a higher rank theory one has to do such transformation
in the entire phase space including ghosts \cite{Hen}.
Things are much easier when one deals with rank one theory where 
there is no need to involve ghosts \cite{Hw}.
In both cases one ends up with constraints
$$p_a = 0. $$
The most general term in the Hamiltonian which is compatible with the
rank of the theory, the class of the constraints and is with non vanishing
contribution to the equation of motion of $q_a$ is
$$
 p_a V_{ab}q_b,
$$
Here $V$ is independent of the dynamical variables but could involve differential
operators in spatial variables if we deal with field theory.
The equations of motion for $q_a$ are
\be
\dot q_a -\lambda_a - (V q)_a=0. \label{em}
\ee
We are looking for a particular solution in the form
$$q_a=e^{V t} \eta_a(t)$$
which allows us easily to find the general solution of eqs.(\ref{em}):
\be
q_a = \bar q_a + \int^t {\rm d}\tau e^{- V \tau} \lambda.  \label{sol}
\ee
Here $\bar q_a$ is the general solutions of the homogeneous part 
of the system (\ref{em}).
It is obvious that $q_a$ could never be zero,
except in the case $\lambda_a =0$ for every $a$.
Thus there are no conditions of the type (\ref{gc}) for any choice
$\lambda\ne 0$.
A possibility to get rid of this contradiction is to use 
gauge conditions with explicit time dependence instead of (\ref{gc}), i.e.
\be
\chi_a(q,p,t) = 0. \label{tgc}
\ee
Another opportunity is to make $\chi_a$ functions of all variables including
$\lambda$ and/or $\pi$
\be
\chi_a(q,p,\lambda, \pi) = 0 \label{mix}
\ee

Let us illustrate the above considerations with a simple example ---
Electrodynamics. 
Common notations for this model are: 
$A_i(x)$ are the configuration space variables,
$E_i(x)$ are the corresponding momenta,
$A_0(x)$ are the Lagrange multipliers,
and $\partial_i E_i =0 $ are the constraints.
A standard gauge of the type (\ref{gc}) is the Columb one ---
$\partial_i A_i =0 $.
Note that this condition does not
fix gauge freedom entirely --- arbitray function of $t$
is a zero mode of $\Delta$.
The Hamiltonian gauge $A_0 = 0$ is of the type (\ref{lc}).
If we use this gauge the ungauged transformations have as parameter
an arbitrary function of the spatial coordinates.
Note that with respect to the residual gauge freedom the two gauge
conditions mentioned above are complementary. 
Using the total constraints (\ref{tot}), which now read
\be
\hat\varphi_a=\pt\pi_a +\partial_i E_i 
\ee
and the Lorenz gauge $\partial_\mu A_\mu$, which is of the type (\ref{mix})
we get $det \Delta \ne 0$ and full gauge fixing.

Now we are ready to construct the BRST charge $Q$. 
Everything is quite standard, except that we shall need extra ghosts
at a particular point.
 Let $c_a$ and $\gP_a$ are the ghost variables, $\{c_a,\gP_b\}=-\delta_{ab}$,
 $c_a$ are real and $\gP_a$ are imagenary. 
The theory is of order one, so for the BRST charge we get \cite{Hen}
\bea
Q'&=& c_a \hat\varphi_a + \half c_a c_b C_{abc}\gP_c =\nn\\
  &=&c_a(\varphi_a - U_{ab} \pi_b + C_{abc}\lambda_b\pi_c) + 
\half c_a c_b C_{abc}\gP_c  + \dot c_a \pi_a. \label{b1}
\eea
There is little use of this expresion because of  the
term $\dot c_a \pi_a$ whose
Poisson bracket with other quantities we cannot calculate.
So, we introduce another set of ghost--antighost pairs 
$\{\bar c_a, \Pg \}$,
substitute $\dot c_a$ in (\ref{b1}) with $i\Pg_a$,
and choose such gauge fixing conditions as to ensure the following
equation of motion
\be
\Pg_a =- i\dot c_a. \label{con}
\ee
The BRST charge now reads
\be
Q''= c_a(\varphi_a - U_{ab} \pi_b + C_{abc}\lambda_b\pi_c) +
 \half c_a c_b C_{abc}\gP_c  + i\Pg_a \pi_a. \label{b2}
\ee
It differs from the BRST charge in the BFV approach  \cite{Hen}
by the  terms $-c_a U_{ab} \pi_b$ and $c_a C_{abc}\lambda_b\pi_c$.
However, the BRST charge thus constructed is not nilpotent. 
We need an additional ghost term to ensure $\{Q,Q\}=0$.
The BRST charge we finally found is:
\be
Q=c_a(\varphi_a - U_{ab} \pi_b + C_{abc}\lambda_b\pi_c + 
C_{abc}\Pg_b\bar c_c) + 
\half c_a c_b C_{abc}\gP_c  + i\Pg_a \pi_a  \label{b3}
\ee
and the BRST invariant Hamiltonian  is
\be
H' = H + c_a U_{ab} \gP_b +\lambda_a U_{ab} \pi_b  
+\Pg_a U_{ab} \bar c_b. \label{h2}
\ee
The difference between (\ref{h2}) and the Hamiltonian in the 
BFV formalism is in the last two terms.

The gauge condition we choose according to
our previous considerations is
\be
\psi=i\bar c_a \chi_a + \gP_a \lambda_a, \label{g3}
\ee
where $\chi_a$ are functions of $q$ and $p$ only.
This choice of $\psi$ coincides with the basic BFV one.
 
For the BRST invariant Lagrangian
$L= \dot q p + \dot\lambda \pi + \dot c \gP + \dot{\bar c}\Pg - H +
\{\psi, Q\} $ 
we obtain
\be
L'= \dot q p + \dot\lambda \pi + \dot c \gP + \dot{\bar c}\Pg -
H - \Pg U \bar c + i\bar c \Delta c + \pi\chi -\lambda\varphi  -
\lambda C \Pg \bar c + i\gP\Pg  + i c C \chi \bar c. \label{brstlag}
\ee
Note that first, the variation of $L'$ with respect to $\gP$ gives 
eqs.(\ref{con}), so our gauge is correct.
Second, due to the fact that $L'$ is quadratic with respect to
ghost momenta it is easy to perform functional integration over them.

It is convenient for comparision to write down also
the corresponding Lagrangian in the BFV approach:
\be
L^{BFV}= \dot q p + \dot\lambda \pi + \dot c \gP + \dot{\bar c}\Pg -
H - c U \gP + i\bar c \Delta c + 
 \pi\chi -\lambda\varphi  - \lambda C c \gP  + i\gP\Pg. \label{bfvlag}
\ee
The only significant difference between $L'$ and $L^{BFV}$ 
 is in the ghost term $i c_a C_{abc}\chi_b\bar c_c$ in (\ref{brstlag}).

\section*{Acknowledgements}

This work is supported by the
Bulgarian National Science Foundation, Grant Ph-1010/00.


\end{document}